\documentclass[aps,prl,twocolumn]{revtex4-1}
\usepackage{amsmath}
\usepackage{amssymb}
\usepackage{graphicx}
\usepackage{bm}

\begin{document}

\title{Generating Two-dimensional Ferromagnetic Charge Density Waves via External Fields}

\author{Heng Jin}
\affiliation{Beijing Computational Science Research Center, Beijing 100193, China}

\author{Jiabin Chen}
\affiliation{Beijing Computational Science Research Center, Beijing 100193, China}

\author{Yang Li}
\affiliation{Beijing Computational Science Research Center, Beijing 100193, China}

\author{Bin Shao}
\affiliation{College of Electronic Information and Optical Engineering, Nankai University, Tianjin 300350, China}

\author{Bing Huang}
\email{bing.huang@csrc.ac.cn}
\affiliation{Beijing Computational Science Research Center, Beijing 100193, China}
\affiliation{Department of Physics, Beijing Normal University, Beijing 100875, China}

\begin{abstract}
Two-dimensional (2D) ferromagnetic charge density wave (CDW), an exotic quantum state for exploring the intertwining effect between correlated charge and spin orders in 2D limit, has not been discovered in the experiments yet. Here, we propose a feasible strategy to realize 2D ferromagnetic CDWs under external fields, which is demonstrated in monolayer VSe$_2$ using first-principles calculations. Under external tensile strain, two novel ferromagnetic CDWs ($\sqrt{3}$$\times$$\sqrt{3}$ and 2$\times2\sqrt{3}$ CDWs) can be generated, accompanied by distinguishable lattice reconstructions of magnetic V atoms. Remarkably, because the driving forces for generating these two ferromagnetic CDWs are strongly spin-dependent, fundamentally different from that in conventional CDWs, the $\sqrt{3}$$\times$$\sqrt{3}$ and 2$\times2\sqrt{3}$ CDWs can exhibit two dramatically different half-metallic phases under a large strain range, along with either a flat band or a Dirac cone around Fermi level. Our proposed strategy and material demonstration may open a door to generate and manipulate correlation effect between collective charge and spin orders via external fields.
\end{abstract}

\maketitle

\draft
\vspace{2mm}
The reduced dimensionality in two-dimensional (2D) materials leads to enhanced correlation effects, which is beneficial for generating multiple symmetry-breaking orders. Of particular interests are the coexistence or competition between charge density wave (CDW) with superconductivity and magnetism in 2D limit \cite{Kiss2007, Ugeda2015, Manzeli2017, Zheng2017}. While the coexistence of CDWs and superconductivity has been observed in some 2D transition-metal dichalcogenides (TMDs) \cite{Kiss2007, Ugeda2015} and consequently stimulated intensive attentions in many other correlated systems \cite{Neupert2021, Mielke2022}, the coexistence of CDWs and ferromagnetism in 2D systems, forming 2D ferromagnetic CDWs, has not been experimentally observed yet. Fundamentally, these unique time-reversal symmetry-breaking CDWs themselves or in combination with superconductivity and topology may contribute to a series of long-sought physics phenomena, e.g., unconventional anomalous Hall effect, chiral charge order, and Kondo lattice \cite{Yang2020, Neupert2021, Vano2021, Mielke2022}.  Practically, the 100\% spin-polarized ferromagnetic CDW, i.e., half-metallic CDW, could be an ideal platform to realize novel CDW-controlled metal-insulator transition in single spin channel for exotic spintronics and information storages \cite{Zutic2004}, beyond the conventional nonmagnetic CDWs \cite{Yoshida2015}. 

The realization of 2D ferromagnetic CDWs is tacitly accepted to be unlikely, mainly due to the mutually exlusive energy gain accounting for the formation of CDW and ferromagnetism: (i) forming CDWs usually reduces the density of states (DOS) at Fermi level ($E_{\rm F}$), accordingly decreasing possibility for forming Stoner itinerant ferromagnetism; (ii) forming ferromagnetism causes splitting of energy bands around $E_{\rm F}$, reshaping topology of Fermi surface (FS) and weakening the condition for FS nesting (FSN), one possible origin of CDWs \cite{Chan1973, Rossnagel2011}. As a typical example, monolayer (ML) VSe$_2$ is able to exhibit multiple CDW phases \cite{Zhang2017, Chen2018, Feng2018, Duvjir2018, Fumega2019, Coelho2019}, room-temperature magnet \cite{Bonilla2018, Yu2019}, and even possible superconductivity \cite{Yilmaz2021}. However, after more than three decades of studies \cite{Zhang2017, Chen2018, Feng2018, Duvjir2018, Fumega2019, Coelho2019, Bruggen1976, Bonilla2018, Yu2019, Manzeli2017, Chua2021, Yilmaz2021}, ferromagnetic CDWs have never been observed in VSe$_2$. A similar situation is also found in other 2D TMD systems \cite{Kiss2007, Ugeda2015, Manzeli2017}. Until now, a feasible strategy to realize 2D ferromagnetic CDW is still lacking, preventing the understanding of new correlation effects between collective charge and spin orders induced novel physics phenomena in monolayer limit.

\textbf{A Strategy for Generating Ferromagnetic CDW.} Here, we propose a feasible strategy to realize ferromagnetic CDW under external field, following by two possible criteria: (i) the system should exhibit ferromagnetic configuration and its competing nonmagnetic CDW as groundstate and metastable state, respectively, making it a possible platform for realizing new complex ground states (left panel, Fig. 1); (ii) an external field could induce dynamic instability of this ferromagnetic configuration but maintains its total energy lower than that of nonmagnetic CDW; the dynamic instability increases as the external field increases. Once criteria (i)-(ii) are satisfied, upon certain external field a sufficiently strong dynamical instability may trigger a new CDW coexisting with ferromagnetic order (middle panel, Fig. 1), as long as ferromagnetic order can survive the lattice reconstruction. Importantly, during the formation of ferromagnetic CDW, an additional criterion [criterion (iii)] may be required to form half-metallic CDW, that is, the major driving force for ferromagnetic CDW formation should be strongly spin-dependent. As a result, significantly different band renormalization could occur in different spin channels, providing a key ingredient for forming half-metallicity (right panel, Fig. 1). In practice, the 2D systems are ideal platforms to realize this strategy due to their convenience to be applied with various external fields, e.g., strain \cite{Lee2008,Guinea2009}, electric \cite{Novoselov2004,Zhang2009}, and irradiation \cite{Cho2015,Sun2016} fields.

\begin{figure}[htb]
\includegraphics[width=\linewidth]{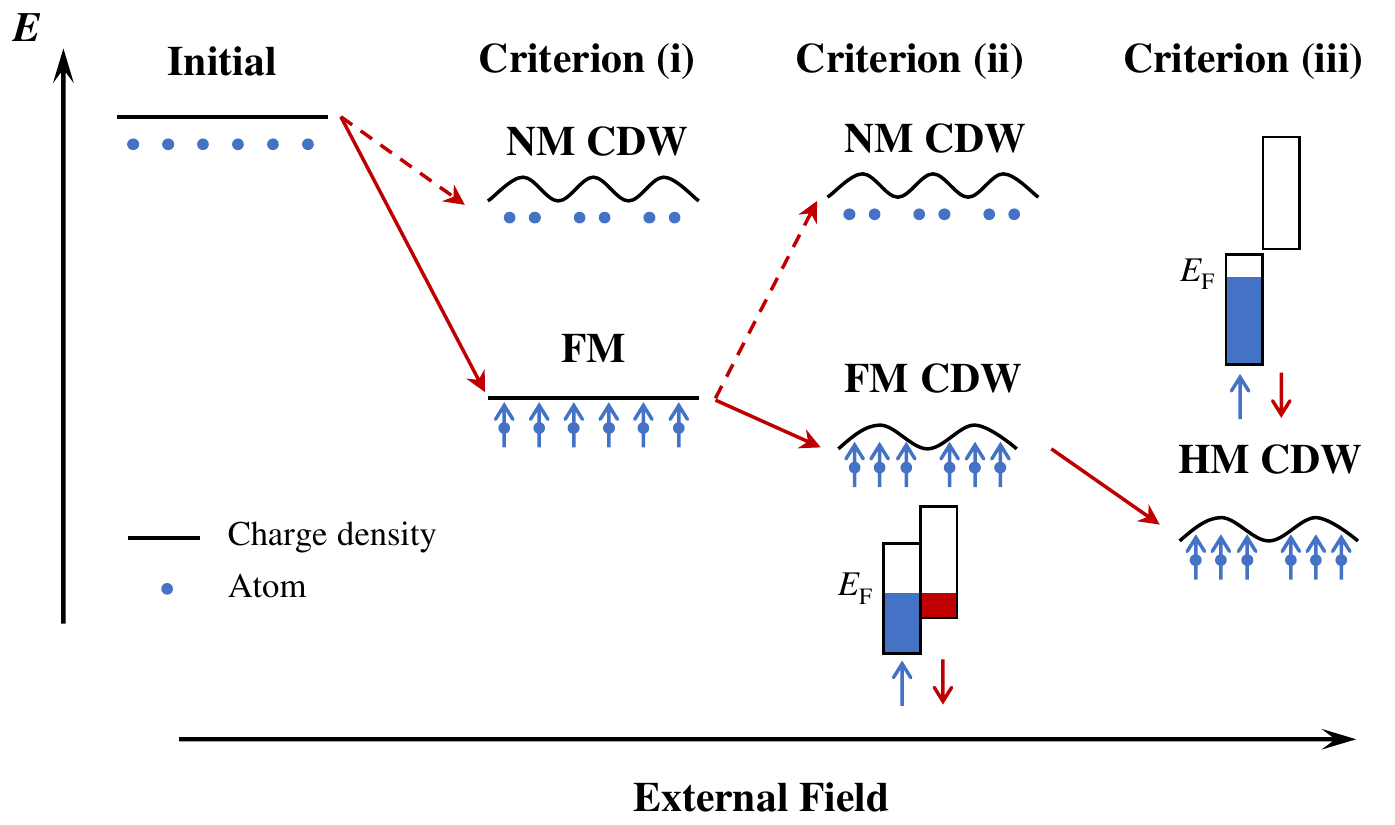}
\caption{A strategy for generating ferromagnetic CDW. Criterion (i) requires the system holding ferromagnetic (FM) phase and nonmagnetic (NM) CDW as the groundstate and metastable state, respectively. Criterion (ii) represents an external field may trigger a new FM CDW formation. Criterion (iii) is an additional requirement for half-metallic (HM) CDW formation. See text for more explanation.}
\end{figure}

In this study, using first-principles calculations (see Methods \cite{Jin2022}), we propose that ML-VSe$_2$ is such an ideal platform to generate ferromagnetic CDWs via our strategy. First, ML-VSe$_2$ holds ferromagnetic 1T phase (nonmagnetic CDWs) as groundstate (metastable states), satisfying criterion (i). Second, under in-plane tensile strain ($\varepsilon$), the total energy of ferromagnetic 1T phase maintains to be lower than that of nonmagnetic CDWs, but its dynamic instability increases as $\varepsilon$ increases, satisfying criterion (ii). Therefore, two ferromagnetic CDW orders, $\sqrt{3}$$\times$$\sqrt{3}$ and 2$\times 2\sqrt{3}$ CDWs, can be generated in ML-VSe$_2$ under certain critical $\varepsilon$. Importantly, the momentum-dependent electron-phonon coupling (MEPC)
and FSN, major driving forces for forming these two ferromagnetic CDWs, are strongly spin-dependent, eventually satisfying criterion (iii). Consequently, the $\sqrt{3}$$\times$$\sqrt{3}$ and 2$\times2\sqrt{3}$ CDWs exhibit
fascinating half-metallic phases under a large range of $\varepsilon$. In particular, the $\sqrt{3}$$\times$$\sqrt{3}$ CDW possesses A-type half-metallic state with a flat band around $E_{\rm F}$, whereas 2$\times2\sqrt{3}$ CDW holds B-type half-metallic state with a clean Dirac cone at $E_{\rm F}$. 

\begin{figure*}[htb]
\includegraphics[width=\linewidth]{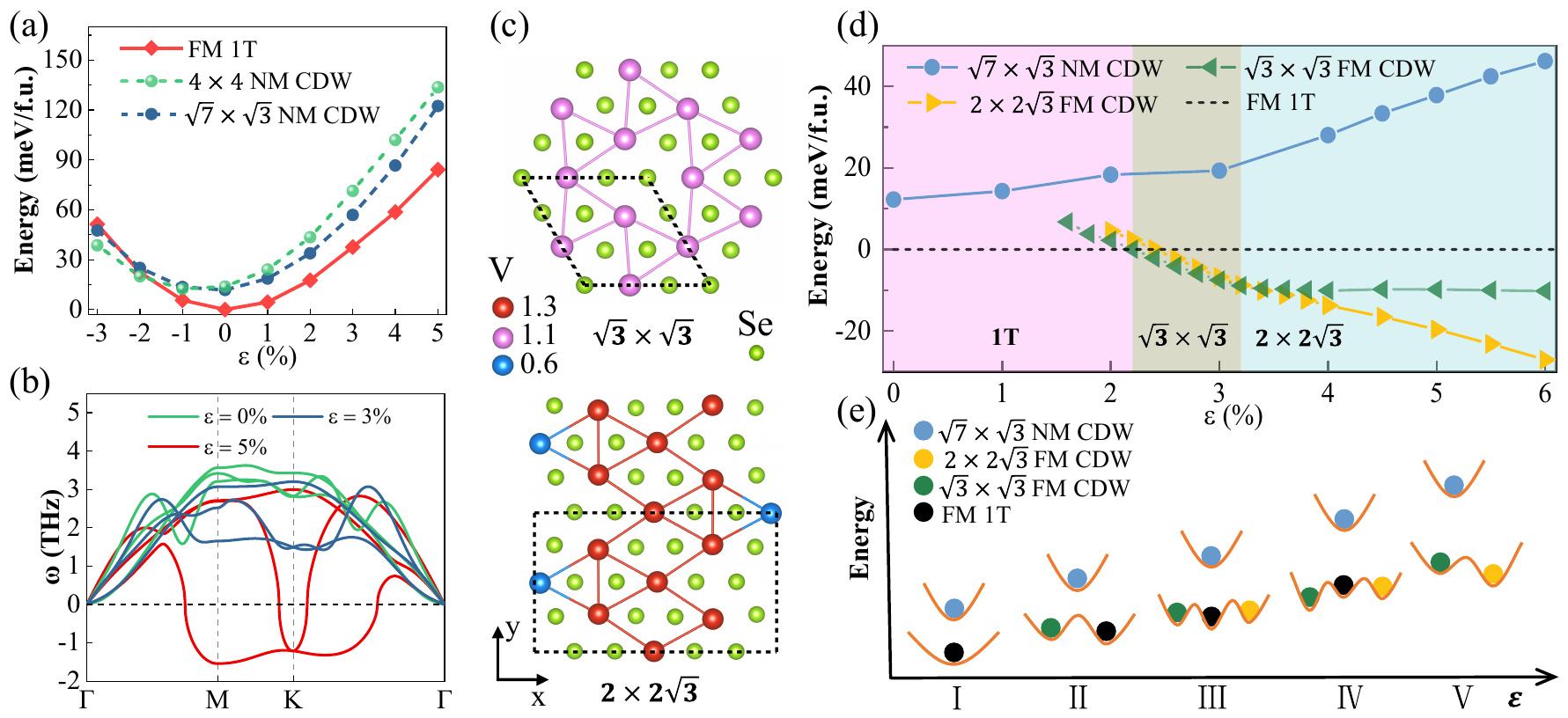}
\caption{Strain-tunable ferromagnetic CDWs in ML-VSe$_2$. (a) Calculated total energies of FM 1T phase and two NM CDWs in ML-VSe$_2$ as a function of strain. (b) Lowest three branches of phonon spectra of FM 1T phase under three typical tensile strain. (c) Top views of $\sqrt{3}$$\times$$\sqrt{3}$ ($\varepsilon$=3\%) and 2$\times2\sqrt{3}$ ($\varepsilon$=5\%) configurations. Primitive cells are marked as dashed-lines. Variable magnetic moments in V atoms are illustrated with different colors (in unit of $\mu_B$). (d) Total energies of FM 1T and various CDW phases as a function of tensile strain. (e) Energy diagram of strain-tunable CDW phase transitions.}
\end{figure*}

\textbf{Generating Ferromagnetic CDWs in ML-VSe$_2$.} The structure of ML 1T-VSe$_2$ has a space group of $P\overline{3}m1$, where one V layer is sandwiched by two Se layers. The calculated lattice constant of ML 1T-VSe$_2$ (3.34 \AA) agrees well with the experimentally measured one ($\sim$3.35 \AA) \cite{Yu2019}. Without external field, ferromagnetic 1T phase is the groundstate, whose total energy is $\sim$24 meV/f.u. lower than that of nonmagnetic one. This ferromagnetic 1T phase has been observed by some experiments at room-temperature \cite{Bonilla2018,Yu2019}. On the other hand, the phonon spectrum of nonmagnetic 1T phase shows strong instability with imaginary phonon modes near 1/2 $\Gamma$-M and 3/5 $\Gamma$-K (Fig. S1 \cite{Jin2022}), indicating the existence of 4$\times$4 and $\sqrt{7}$$\times$$\sqrt{3}$ CDWs (Table S1 \cite{Jin2022}), respectively. Indeed, these nonmagnetic CDWs have also been observed in experiments \cite{Zhang2017,Chen2018,Feng2018,Duvjir2018,Fumega2019,Coelho2019}. Importantly, the calculated total energy of ferromagnetic 1T phase is $\sim$13.8 and $\sim$11.7 meV/f.u. lower than that of nonmagnetic 4$\times$4 and $\sqrt{7}$$\times$$\sqrt{3}$ CDWs, respectively, which indicates that ML-VSe$_2$ meets the criterion (i) for forming ferromagnetic CDWs.

To check whether criterion (ii) can be simultaneously satisfied, we have further calculated the energy and dynamic instability of ferromagnetic 1T phase under external strain field. As shown in Fig. 2(a), the energy difference ($\Delta$E) between ferromagnetic 1T phase and these two nonmagnetic CDWs increases (decreases) as a function of tensile (compressive) $\varepsilon$, indicating that the energetic stability of corresponding ferromagnetic 1T phase can be enhanced under tensile $\varepsilon$. Surprisingly, when tensile $\varepsilon$ is applied, although the total energy of ferromagnetic 1T phase can be further lowered with respect to the nonmagnetic CDWs, its dynamic instability increases, as indicated by the calculated phonon spectra in Fig. 2(b). The larger the $\varepsilon$, the softer the phonon modes. Therefore, ML-VSe$_2$ can also meet the criterion (ii) for forming ferromagnetic CDWs. Particularly, when $\varepsilon$=5\%, two imaginary phonon modes appear, i.e., one is located along M-K-$\Gamma$ and the other is located at K. The maximum at M (K) suggests a 2$\times$2 ($\sqrt{3}$$\times$$\sqrt{3}$) CDW, while maximum near 3/4 $\Gamma$-K suggests a 2$\times2\sqrt{3}$ CDW (Table S1 \cite{Jin2022}).

While it is found that 2$\times$2 CDW is dynamically unstable with a spontaneous transformation to 2$\times2\sqrt{3}$ CDW within a doubled supercell, both 2$\times2\sqrt{3}$ and $\sqrt{3}$$\times$$\sqrt{3}$ CDWs are dynamically stable under tensile strain (Fig. S2 \cite{Jin2022}). Importantly, the ferromagnetic order of both CDWs can survive the formation of CDW. As shown in Fig. 2(c), the $\sqrt{3}$$\times$$\sqrt{3}$ CDW (space group: $P3m1$) is formed by anti-trimerization of V atoms in 1T phase while the 2$\times2\sqrt{3}$ CDW (space group: $P2_1$) is formed by the reconstruction of V atoms into a zigzag stripe. Besides, the local magnetic moments of V atoms are uniformly distributed in $\sqrt{3}$$\times$$\sqrt{3}$ CDW but exhibit a novel spin oscillation along $x$ axis in 2$\times2\sqrt{3}$ CDW [Fig. 2(c)].

In Fig. 2(d), we have calculated the total energies of these ferromagnetic CDWs as a function of tensile $\varepsilon$, compared with ferromagnetic 1T phase and nonmagnetic $\sqrt{7}$$\times$$\sqrt{3}$ CDW. Interestingly, $\varepsilon$-tunable multiple phase transitions between ferromagnetic 1T and ferromagnetic CDWs are observed. When 0$\le$$\varepsilon$$<$1.8\%, the ferromagnetic 1T is groundstate, while ferromagnetic CDWs cannot exist [I in Fig. 2(e)]. When 1.8$\le$$\varepsilon$$<$2\%, $\sqrt{3}$$\times$$\sqrt{3}$ CDW appears as a metastable state [II in Fig. 2(e)]. When 2$\le$$\varepsilon$$<$2.2\%, 2$\times$2$\sqrt{3}$ CDW emerges as another metastable state [III in Fig. 2(e)]. Importantly, when 2.2$\le$$\varepsilon$$<$3.2\%, the first groundstate phase transition occurs from 1T to $\sqrt{3}$$\times$$\sqrt{3}$ CDW [IV in Fig. 2(e)], i.e., the ferromagnetic CDW is now stabilized as the groundstate. When $\varepsilon$$\geq$3.2\%, the second groundstate phase transition occurs from $\sqrt{3}$$\times$$\sqrt{3}$ CDW to 2$\times$2$\sqrt{3}$ CDW; meanwhile, the 1T phase is no longer a metastable state [V in Fig. 2(e)]. On the other hand, the $\Delta$E between nonmagnetic $\sqrt{7}$$\times$$\sqrt{3}$ CDW and these two ferromagnetic CDWs increase as $\varepsilon$ increases, i.e., the larger the $\varepsilon$, the stronger the instability of nonmagnetic CDW.

We have further confirmed the energetic stability of ferromagnetic orders in $\sqrt{3}$$\times$$\sqrt{3}$ and 2$\times2\sqrt{3}$ CDWs by comparing them with other typical antiferromagnetic orders (Fig. S3 and Table. S2 \cite{Jin2022}). It turns out that ferromagnetic orders are always energetically favorable for both $\sqrt{3}$$\times$$\sqrt{3}$ and 2$\times2\sqrt{3}$ CDWs under tensile strain. The larger $\varepsilon$ results in more stable ferromagnetic orders. Interestingly, calculated magnetocrystalline anisotropy energy (MAE) shows that the easy axis of ferromagnetic CDWs is along in-plane direction, similar to CrCl$_3$ \cite{BedoyaPinto2021}; and the larger the $\varepsilon$, the larger the MAE. These results indicate that a larger $\varepsilon$ can induce a higher $T_c$ in these ferromagnetic CDWs.

\begin{figure}[htb]
\includegraphics[width=\linewidth]{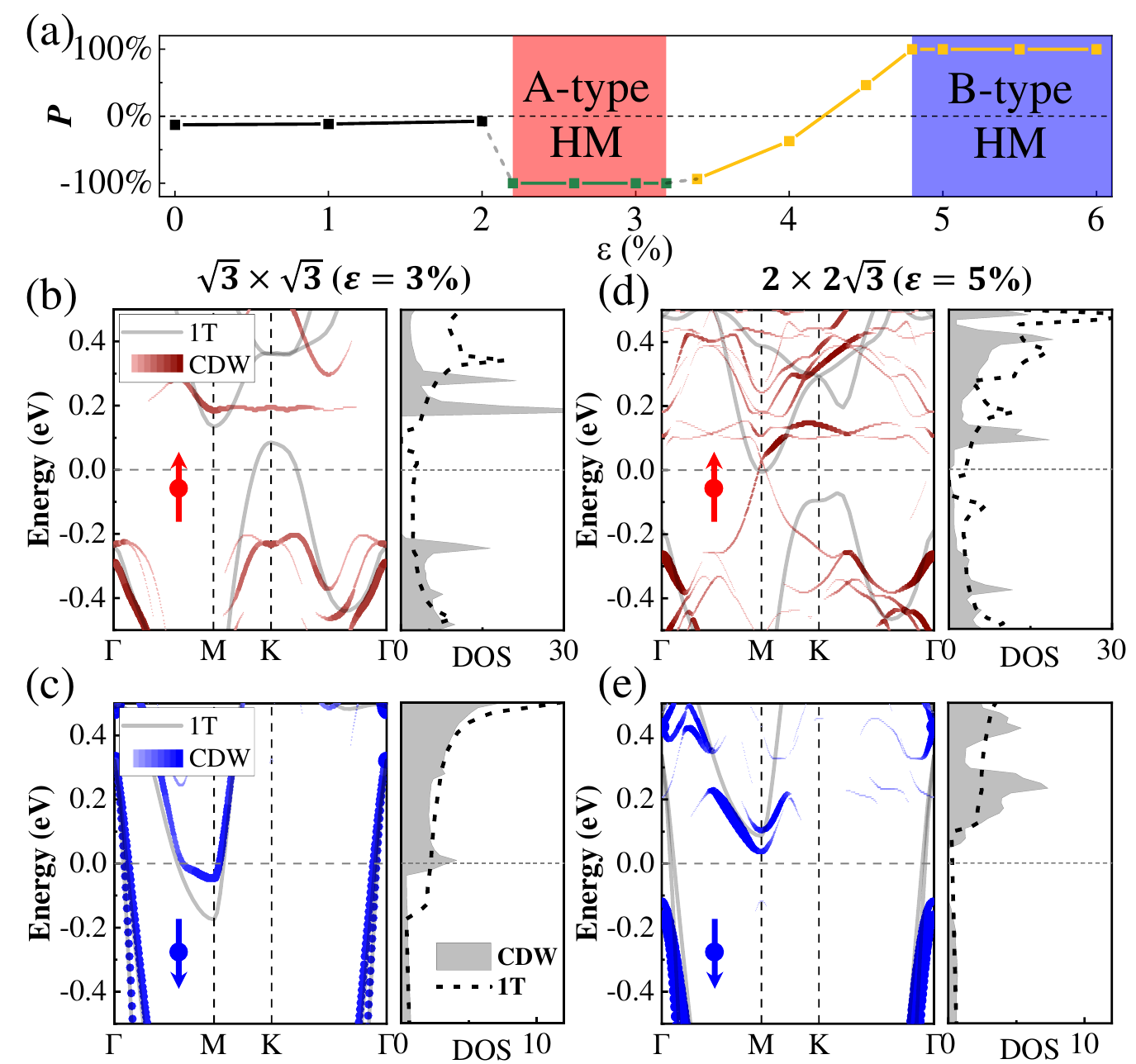}
\caption{Electronic structures of ferromagnetic CDWs. (a) Spin-polarization ratio $P$ as a function of strain for different groundstates. Unfolded band structures and DOS for $\sqrt{3}$$\times$$\sqrt{3}$ CDW ($\varepsilon$=3\%) in (b) spin $\uparrow$ and (c) spin $\downarrow$ channels. (d)-(e): same as (b)-(c) but for 2$\times2\sqrt{3}$ CDW ($\varepsilon$=5\%). Band structures of FM 1T VSe$_2$ are also plotted here for comparison. Fermi level is set to zero.}
\end{figure}

\textbf{Electronic Structures of Ferromagnetic CDWs}. It is important to further understand the electronic structures of these two ferromagnetic CDWs. Here spin-polarization ratio $P$ is defined as $P$=$(N_{\uparrow}(E_{\rm F})-N_{\downarrow}(E_{\rm F}))/(N_{\uparrow}(E_{\rm F})+N_{\downarrow}(E_{\rm F}))$ \cite{Sun2017}, where $N_{\uparrow}(E_{\rm F})$ and $N_{\downarrow}(E_{\rm F})$ are the DOS values at $E_{\rm F}$ in spin $\uparrow$ (majority) and spin $\downarrow$ (minority) channels, respectively. The calculated $\varepsilon$-dependent $P$ for groundstates is summarized in Fig. 3(a). When 0$\le$$\varepsilon$$<$2.2\%, the $P$ of ferromagnetic 1T phase is very small ($\sim$10\%) and insensitive to $\varepsilon$. Remarkably, when 2.2$\leq$$\varepsilon$$<$3.2\%, the $P$ of $\sqrt{3}$$\times$$\sqrt{3}$ CDW keeps to be -100\%, denoted as A-type half-metallicity [majority (minority) spin channel is insulating (metallic)]. When 3.2$\leq$$\varepsilon$$<$4.8\%, the bandgap in spin $\uparrow$ ($\downarrow$) channel of 2$\times2\sqrt{3}$ CDW gradually decreases (increases) (Fig. S4 \cite{Jin2022}), along with $P$ gradually changing from $\sim$-90\% to 100\%. When $\varepsilon$$\geq$4.8\%, the $P$ of 2$\times2\sqrt{3}$ CDW keeps to be 100\%, denoted as B-type half-metallicity [majority (minority) spin channel is conducting (insulating)]. In practice, these two different types of half-metallic phases can be distinguishable using the spin-resolved STM measurements (see Fig. S5 \cite{Jin2022} and related discussion).

In the following, the $\sqrt{3}$$\times$$\sqrt{3}$ ($\varepsilon$=3\%) and 2$\times2\sqrt{3}$ ($\varepsilon$=5\%) CDWs are selected to explore their unusual half-metallic states. For $\sqrt{3}$$\times$$\sqrt{3}$ CDW, in spin $\uparrow$ channel [Fig. 3(b)], the hole pocket centered at K in FM-1T phase is largely suppressed and pushed down to a lower energy position during
the CDW formation, forming a Mexican-Hat-shape band dispersion with a large DOS peak at VBM. Importantly, the strong band renormalization results in a unusual flat band, composed with twofold V $d$ orbitals, in the bottom of conduction band along M-K-$\Gamma$, reflected by the very sharp DOS peak at CBM. The formation of this flat-band might be due to the anti-trimerization of V atoms in a triangular-like frustrated lattice \cite{Calugaru2021}. Furthermore, this flat band can gradually shift down to $E_{\rm F}$ under larger tensile $\varepsilon$ and electron doping (Fig. S6 \cite{Jin2022}). In spin $\downarrow$ channel [Fig. 3(c)], the energy bands are much less changed during the CDW formation, i.e., the electron pocket centered at M is slightly pushed up. Eventually, an A-type half-metallic phase is formed. Interestingly, DOS at $E_{\rm F}$ in spin $\uparrow$ ($\downarrow$) channel is fully suppressed (remarkably increased) during the $\sqrt{3}$$\times$$\sqrt{3}$ CDW formation, invaliding the common expectation that CDW formation is accompanied by DOS reduction at $E_{\rm F}$ \cite{Lian2018,Tan2021,Seong2021}. This highly asymmetrical band renormalization in spin $\uparrow$ and $\downarrow$ channels result in a large half-metallic bandgap $\sim$0.4 eV.

Figs. 3(d) and 3(e) shows the calculated spin $\uparrow$ and $\downarrow$ band structures of 2$\times 2\sqrt{3}$ CDW, respectively. A larger band renormalization also occurs in spin $\uparrow$ channel than in spin $\downarrow$ channel during CDW formation, along with largely-reduced (fully-suppressed) DOS at $E_{\rm F}$ in spin $\uparrow$ ($\downarrow$) channel. Accordingly, a B-type half-metallic phase with insulating state located in spin $\downarrow$ channel
is formed in 2$\times2\sqrt{3}$ CDW. Unexpectedly, as shown in Fig. 3(d), a clean Dirac cone located at M around $E_{\rm F}$ appears in the unfolded band structure of spin $\uparrow$ channel, accompanied by the suppression of local bandgap during CDW formation, which may be visible under spin-resolved ARPES measurements \cite{Lv2019}. In spin $\downarrow$ channel [Fig. 3(e)], the hole pocket centered at $\Gamma$ is strongly pushed down by $\sim$0.4 eV, which not only plays a key role in forming a half-metallic bandgap $\sim$0.2 eV, but also introduces a new valley state around $\Gamma$.

\begin{figure}[htb]
\includegraphics[width=\linewidth]{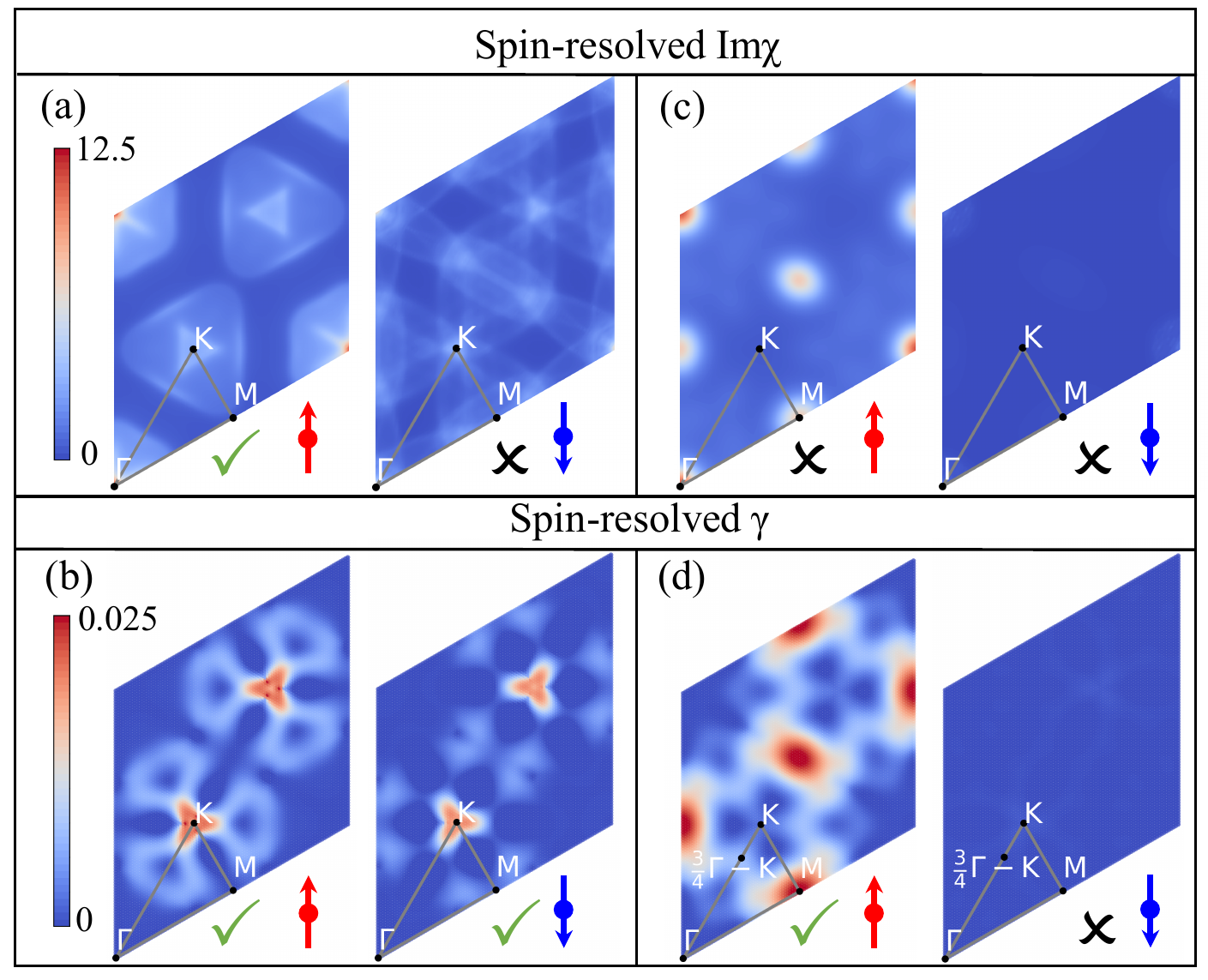}
\caption{Origins of strain-dependent ferromagnetic CDWs. (a) Spin-resolved Im$\chi$ and (b) $\gamma$ (contributed by the lowest phonon mode in the phonon spectrum) for FM 1T-VSe$_2$ under $\varepsilon$=3\%. (c)-(d): same as (a)-(b) but under $\varepsilon$=5\%. Unit of $\gamma$ is THz.}
\end{figure}

\textbf{Origins of Ferromagnetic CDWs under Strain}. The driving force behind the CDW transition is an important subject with ongoing controversy \cite{Johannes2008,Zhu2015,Zheng2017}. The strong spin-dependent band renormalization in $\sqrt{3}$$\times$$\sqrt{3}$ and 2$\times2\sqrt{3}$ CDWs indicate that the driving force for these CDW formation may have a strong spin dependence feature. Generally, FSN and/or MEPC, common origins of the observed CDWs in many TMD systems \cite{Manzeli2017, Rossnagel2011}, are spin-independent. The real (Re$\chi$) and imaginary (Im$\chi$)
parts of electron susceptibility $\chi$ reflect the electron instability and FSN of a system \cite{Johannes2008}, respectively. When both Re$\chi$ and Im$\chi$ peak at the same wave vector $\mathbf{q}$, a FSN-induced CDW with $\mathbf{q}_{\rm {CDW}}$ may be triggered. Furthermore, MEPC can be evaluated by calculating phonon linewidth $\gamma$ \cite{Diego2021}.

Figure 4 shows the calculated spin-resolved Im$\chi$ and $\gamma$ of ferromagnetic 1T-VSe$_2$ under two typical $\varepsilon$, which indeed exhibit strong spin dependence, satisfying criterion (iii) for forming half-metallic CDWs. For $\varepsilon$=3\%, as shown in Fig. 4(a) and Fig. S7 \cite{Jin2022} , noticeable peaks appear around K in both Re$\chi$ and Im$\chi$ of spin $\uparrow$ channel, but not of spin $\downarrow$ channel. Contributed by hole pocket round K [Fig. 3(b)], FSN of spin $\uparrow$ channel rather than spin $\downarrow$ channel could contribute to the formation of half-metallic $\sqrt{3}$$\times$$\sqrt{3}$ CDW (Table S1 \cite{Jin2022}). In addition, it may also provide a valid understanding for the much
larger band renormalization in spin $\uparrow$ [Fig. 3(b)] than spin $\downarrow$ [Fig. 3(c)] channel, a key factor for forming half-metallicity. Meanwhile, as shown in Fig. 4(b), high intensity peaks of $\gamma$ appear around K in both spin channels, suggesting that MEPC may be another possible reason to form half-metallic $\sqrt{3}$$\times$$\sqrt{3}$ CDW. Therefore, the joint MEPC and (spin-dependent) FSN may contribute to the half-metallic $\sqrt{3}$$\times$$\sqrt{3}$ CDW formation.

The situation is different for $\varepsilon$=5\%. A hole pocket around K [Fig. 3(b)] of spin $\uparrow$ channel in ferromagnetic 1T phase under $\varepsilon=3\%$ is now converted to electron pocket at M at FS [Fig. 3(d)], changing the FSN condition dramatically. Indeed, as shown in Fig. 4(c), the peak in Im$\chi$ shifts from K to M in spin $\uparrow$ channel, which, unfortunately, cannot contribute to the $2\times 2\sqrt{3}$ CDW formation (Table S1 \cite{Jin2022}). In the spectra of $\gamma$ [Fig. 4(d)], we can observe three distinctive peaks around M, 3/4 $\Gamma$-K and K in spin $\uparrow$ channel. The peak at 3/4 $\Gamma$-K in spin $\uparrow$ channel indicates that the key driving force to form half-metallic 2$\times 2\sqrt{3}$ CDW may be MEPC. Meanwhile, no noticeable peaks are observed in both Im$\chi$ and $\gamma$ in spin $\downarrow$ channel. This may account for the observed larger band renormalization in spin $\uparrow$ [Fig. 3(d)] than spin $\downarrow$ [Fig. 3(e)] channel. Hence, the spin-selective MEPC, rather than FSN, may play a major role in forming half-metallic 2$\times 2\sqrt{3}$ CDW. Figure 4 reveals the unusual relationship between CDW transition and half-metallicity formation.

\textbf{Outlook}. In the current experiments, some popular substrates with smaller lattice-constants than ML-VSe$_2$ are frequently applied for growing ML-VSe$_2$ \cite{Bonilla2018, Coelho2019, Feng2018, Fumega2019}. It might induce some compressive $\varepsilon$ to ML-VSe$_2$, effectively suppressing the energetic stability of ferromagnetic order and even stabilizing the nonmagnetic CDWs as groundstates [Fig. 2(a)]. To meet the criteria to form ferromagnetic CDWs, we suggest that the ML-VSe$_2$ should be prepared on the substrates with larger lattice-constant, or directly apply tensile strain, e.g., via bending \cite{Han2019}. In general, our strategy for generating ferromagnetic CDWs are valid in many other 2D systems under various external fields, not limited to strain field.

\textbf{Acknowledgement}. The authors thank Prof. Z. Liu for helpful discussions. This work is supported by the NSFC (Grant No. 12088101) and NSAF (Grant No. U1930402). Computations are done at Tianhe-JK supercomputer at CSRC.

\bibliographystyle{apsrev4-2}
\nocite{*}
%
\end{document}